\documentclass[amsmath,prb,twocolumn,showpacs,superscriptaddress]{revtex4}

\usepackage{color}
\usepackage{amsfonts}
\usepackage{graphicx}

\begin{document}

\title{Molecular nanoplasmonics: self-consistent electrodynamics in current carrying junctions}
\author{Alexander J. White}
\email{ajw009@ucsd.edu}
\affiliation{Department of Chemistry \& Biochemistry, University of California at San Diego, La Jolla, CA 92093, USA}
\author{Maxim Sukharev}
\email{maxim.sukharev@asu.edu}
\affiliation{Department of Applied Sciences and Mathematics, Arizona State University, Mesa, AZ 85212, USA}
\author{Michael Galperin}
\email{migalperin@ucsd.edu}
\affiliation{Department of Chemistry \& Biochemistry, University of California at San Diego, La Jolla, CA 92093, USA}

\begin{abstract}
We consider a biased molecular junction subjected to external time-dependent
electromagnetic field. 
We discuss local field formation due to both surface plasmon-polariton excitations
in the contacts and the molecular response.
Employing realistic parameters we demonstrate that such self-consistent treatment
is crucial for proper description of the junction transport characteristics.
\end{abstract}


\maketitle
\section{Introduction}\label{intro}
 Research in plasmonics is expanding its domains into several sub-fields due to significant advances in experimental techniques.
 \cite{Hutter:2004ae,Maier:2005zp,Zayats:2005gl,Barnes:2007lr,HalasNL10,Stockman:2011bw}
The unique optical properties of the surface plasmon-polariton (SPP) resonance, being the very foundation of plasmonics, find intriguing applications in 
optics of nano-materials,\cite{BryantLPR08,NoginovNATURE09,GeissenNatMat09}
materials with effective negative index of refraction,\cite{Sarychev:2007lu,QuidantDrachevOME11,HessNatMat12}
direct visualization,\cite{StockmanNJP08,ZewailNL12}
photovoltaics,\cite{Catchpole:2008oq,PorsOC11,RichardsSPIE12}
single molecule manipulation,\cite{ReuterSukharevSeidemanPRL08,MayerSMALL10,SeidemanNL10}
and biotechnology.\cite{Homola:2003hf,DelceaACSNano12,SannomiyaVorosTB11,SintonAC09}
Theoretical modeling of optical properties of metal nanostructures 
is conventionally based on numerical integration of Maxwell's equations,\cite{Gray:2003ub,YelkSukharevSeidemanJCP08,LopataNeuhauserJCP11,SchatzVanDuyneJPCC11,SchatzJPCC12} 
although simulations within time-dependent density functional
theory appeared recently for small atomic clusters.\cite{NordlanderNL09,JensenCR11} Moreover, current theoretical models are quickly advancing toward self-consistent simulations of hybrid materials: metal/semiconductor nanostructures optically coupled to ensembles of quantum emitters.\cite{SukharevNitzanPRA11} 
This methodology, based on numerical integration of corresponding Maxwell-Bloch equations, 
brings new insights into nano-optics as it allows for the capture of collective effects.

The molecular optical response in a close proximity of plasmonic materials is greatly enhanced 
by SPP modes leading to the discovery of the single molecule spectroscopy.\cite{VanDuyne1977,NieEmoryScience97,ZhangNL07}
Recently, experiments performed on current carrying molecular 
junctions started to appear.\cite{CheshnovskySelzerNatNano08,NatelsonNL08,HoPRB08,AllaraNL10,NatelsonNN11}
Theoretical modeling of molecule-SPP systems utilizes the tools of quantum 
mechanics for the molecular part. In particular, studies of optical response 
of isolated molecules absorbed on metallic nanoparticles utilize
Maxwell-Bloch (Maxwell-Schr\"{o}dinger)\cite{SukharevNitzanPRA11,LopataNeuhauserJCP09,NeuhauserJCP09,NeuhauserJCP11,PriorSeidemanSukharevPRL12} equations
or near field-time dependent density functional theory formulations.\cite{GaoNeuhauserJCP12,SchatzJPCA12}

Realistic molecular devices are open quantum systems exchanging 
energy and electrons with surrounding environment (baths). 
This is especially important in studies of molecules in current carrying
junctions interacting with external fields.\cite{GalperinNitzanPCCP12} 
Usually in such studies the electromagnetic (EM) field is assumed to be 
an external driving force.\cite{FainbergNitzanRatnerNL12,PeskinGalperinJCP12,ParkGalperinEPL11,ParkGalperinPRB11,MayJPCC10,FainbergPRB10,GalperinRatnerNitzanJCP09,CuevasPRB08,FainbergNitzanPRB07,CuevasPRB07,GalperinNitzanJCP06,KohlerLehmannHanggiPR05} 
Recently we utilized the nonequilibrium Green function technique to study
the transport and optical response of a molecular junction subjected to external
EM field taking into account near-fields driven by SPP local modes, specific for a particular junction geometry.\cite{Sukharev2010,FainbergPRB11} 
Although the formulation allows us to describe the molecular junction 
with formation of the local field by SPP excitations in the contacts
taken into account explicitly, the molecular influence on formation of the local EM
field was disregarded in these studies. 
Note that such influence was shown to have measurable effects in plasmonic spectrum.\cite{PriorSeidemanSukharevPRL12,SukharevNitzanPRA11,NordlanderNatMater10,LopataNeuhauserJCP09} 

When a molecule located near metal surface is driven by a strong EM field, one can expect to observe significant changes in the total EM field due to radiation emitted by the molecule. Such radiation 
although quickly degrading with the distance from molecular position can nevertheless 
noticeably alter the local EM field. Since the latter is driving the molecule, transport characteristics of 
the junction may be significantly modified.  This calls for a self-consistent treatment,
where both SPP excitations and molecular response participate in formation of the local EM field. 

Here we extend our previous considerations by taking into account complete electrodynamics
and molecular junction response in a self-consistent 
manner combining Maxwell's equations with electron transport dynamics. 
The molecule is treated as a pointwise source in the Ampere law.
We demonstrate the importance of the molecular response in the formation
of the local field for an open molecular system far from equilibrium.
The effect is shown to be important for proper description of the junction transport 
characteristics. The paper is organized as follows. 
Section \ref{model} presents a transport model of the molecular junction.
Section \ref{EM} describes the methodology of computing the EM field taking into account
molecular response. The results are presented in section \ref{numres}.
Section \ref{conclude} summarizes our work.

\section{\label{model}Molecular junction subjected to external EM field}
\begin{figure}[t]
\centering\includegraphics[width=\linewidth]{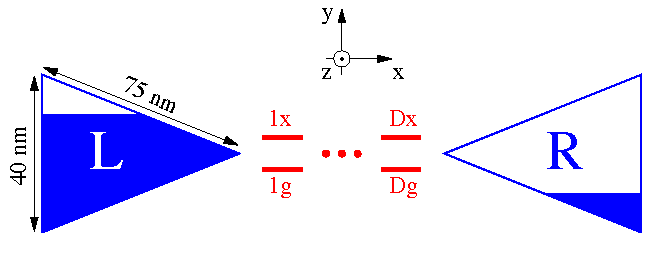}
\caption{\label{fig1}
(Color online) A sketch of the junction.
}
\end{figure}

We consider a junction with a molecular bridge ($M$) connecting between two
contacts ($L$ and $R$). The bridge is formed by $D$ two-level systems with 
the levels representing ground ($g$) and excited ($x$) states of the molecule. 
Each of the two level systems is subjected to a classical local EM field $\vec E(t)$
(see section~\ref{EM} for details of its calculation). 
Electron transfer is allowed along the chain of ground (excited) levels of the bridge. 
The contacts are taken in the form of bowtie antennas, and are assumed to be
reservoirs of free electrons each in its own equilibrium 
with electrochemical potentials $\mu_L$ and $\mu_R$, respectively
(see Fig.~\ref{fig1}).
The Hamiltonian of the system reads (here and below $e=\hbar=1$)
\begin{align}
 \label{H}
 \hat H(t) =& \hat H_M(t) + \sum_{K=L,R}\left(\hat H_K + \hat V_K\right)
 \\
 \label{HM}
 \hat H_M(t) =& \sum_{s=g,x} \left[
    \sum_{m=1}^D\varepsilon_{s}\hat d_{ms}^\dagger\hat d_{ms}
    -\sum_{m=1}^{D-1} t_s\left(\hat d_{m+1s}^\dagger\hat d_{ms}+H.c.\right)
    \right]
 \nonumber \\
     -& \sum_{m=1}^D\left(\vec\mu_{mg,mx}\hat d_{mg}^\dagger\hat d_{mx}
         +H.c.\right)\vec E_m(t)    
\\
 \label{HK}
 \hat H_K =& \sum_{k\in K} \varepsilon_k \hat c_k^\dagger\hat c_k
\\
 \label{VK}
 \hat V_K =& \sum_{k\in K}\sum_{s=g,x}\left(V_{k,m_Ks}\hat c_k^\dagger\hat d_{m_Ks}
                 +H.c\right)
\end{align}
where $\hat H_M(t)$ and $\hat H_K$ are Hamiltonians of the molecular bridge ($M$) 
and the contacts ($K=L,R$), and $\hat V_K$ is coupling between them.
In Eqs.~(\ref{HM})-(\ref{VK}) $\hat d_{ms}^\dagger$ ($\hat d_{ms}$) and 
$\hat c_k^\dagger$ ($\hat c_k$) are creation (annihilation) operators for an electron on 
the level $s$ of the molecular bridge site $m$ and state $k$ of the contact, respectively. 
$\vec E_m(t)$ is the local time-dependent field at bridge site $m$, 
and $\vec\mu_{ms,ms'}=\langle ms\rvert\hat{\vec\mu}\lvert ms'\rangle$ is the matrix element of 
the transition molecular (vector) dipole operator between states $\lvert ms\rangle$ and $\lvert ms'\rangle$. 
For simplicity below we assume that the transition dipole moment is the same for 
all bridge sites and has only one non-zero component, $\mu_{mg,mx}\equiv\mu_{gx}$ for any $m$.
$t_s$ ($s=g,x$) and $V_{k,m_Ks}$
are matrix elements for electron transfer in the molecular bridge and between
molecule and contacts, respectively, and $m_K=1$ ($D$) for $K=L$ ($R$).
Note that treating the external field classically allows us to account for arbitrary time dependence
exactly (i.e. beyond perturbation theory).\cite{Sukharev2010} 

We follow the formulation of Ref.~\onlinecite{Sukharev2010}.
Time-dependent current at interface $K$ ($L$ or $R$) is\cite{JauhoWingreenMeirPRB94}
\begin{equation}
 \label{IK}
 I_K(t)=-\mbox{Im}\,\mbox{Tr}\left[
 \mathbf{\Gamma}^K\left(\mathbf{G}^{<}(t,t)
 +\int\frac{d\epsilon}{\pi}\, f_K(\epsilon)\,\mathbf{G}^r(t,\epsilon)\right)
 \right]
\end{equation}
where $\mbox{Tr}[\ldots]$ is a trace over the molecular subspace,
$f_K(\epsilon)\equiv\left[e^{(\epsilon-\mu_K)/T}+1\right]^{-1}$ is the Fermi-Dirac distribution
in contact $K$, $\mathbf{\Gamma}^K$ is the molecular dissipation matrix due to coupling
to contact $K$
\begin{equation}
\label{GammaK}
 \Gamma^K_{m_1s_1,m_2s_2}(\epsilon)\equiv 2\pi\sum_{k\in K}V_{m_1s_1,k}V_{k,m_2s_2}
 \delta(\epsilon-\varepsilon_k),
\end{equation}
and $\mathbf{G}^{<(r)}$ is a matrix in the molecular basis of 
the lesser (retarded) projection of the single particle Green function,
defined on the Keldysh contour as\cite{HaugJauho_2008}
\begin{equation}
\label{G}
 G_{m_1s_1,m_2s_2}(\tau_1,\tau_2)\equiv -i\langle T_c\,
 \hat d_{m_1s_1}(\tau_1)\,\hat d_{m_2s_2}^\dagger(\tau_2)\rangle
\end{equation}
Here $T_c$ is the contour ordering operator and $\tau_{1,2}$ are the contour variables.
In Eq.(\ref{IK}) $\mathbf{G}^r(t,\epsilon)$ is the right Fourier transform of the retarded projection
of the Green function (\ref{G})
\begin{equation}
\label{GrtE}
 \mathbf{G}^r(t,\epsilon)\equiv \int dt'\, e^{i\epsilon(t-t')}\, \mathbf{G}^r(t,t') 
\end{equation}
Note that in Eq.(\ref{IK}) and below we assume the wide band limit\cite{Mahan_1990} in the metallic contacts. 

The Green functions in (\ref{IK}) satisfy the following set of equations of motions\cite{FainbergPRB11,GalperinTretiakJCP08}
\begin{align}
\label{EOMGr}
& i\frac{\partial}{\partial t}\mathbf{G}^r(t,\epsilon) =\mathbf{I}-
\left(\epsilon\mathbf{I}-\mathbf{H}_M(t)+\frac{i}{2}\mathbf{\Gamma}\right)\mathbf{G}^r(t,\epsilon)
\\
\label{EOMGlt}
& i\frac{d}{dt}\mathbf{G}^{<}(t,t) = \left[\mathbf{H}_M(t);\mathbf{G}^{<}(t,t)\right]
-\frac{i}{2}\left\{\mathbf{\Gamma};\mathbf{G}^{<}(t,t)\right\}
\nonumber \\
&+i\sum_{K=L,R}\int\frac{d\epsilon}{2\pi}f_K(\epsilon)
  \left(\mathbf{\Gamma}^K\mathbf{G}^a(\epsilon,t)-\mathbf{G}^r(t,\epsilon)\mathbf{\Gamma}^K\right)
\end{align}
where $\mathbf{I}$ is the unity matrix, $\mathbf{H}_M(t)$ is a representation of the
operator (\ref{HM}) in the molecular basis, 
$\mathbf{\Gamma}\equiv\sum_{K=L,R}\mathbf{\Gamma}^K$,
$[\ldots;\ldots]$ and $\{\ldots;\ldots\}$ are the commutator and anti-commutator,
and $\mathbf{G}^a(\epsilon,t)\equiv[\mathbf{G}^r(t,\epsilon)]^\dagger$.
The first order differential equations (\ref{EOMGr}) and (\ref{EOMGlt}) are solved starting from 
the initial condition of the biased junction steady-state in the absence of the optical pulse,
$E(t=0)=0$
\begin{align}
\label{ICGr}
\mathbf{G}^{r}_0(\epsilon)\equiv&\mathbf{G}^{r}(t=0,\epsilon)
 = \left[\epsilon\mathbf{I}-\mathbf{H}_M(t=0)+\frac{i}{2}\mathbf{\Gamma}\right]^{-1}
\\
\label{ICGlt}
\mathbf{G}^{<}_0\equiv&\mathbf{G}^{<}(t=0,t=0)
\nonumber \\
 =& i\sum_{K=L,R}\int\frac{d\epsilon}{2\pi}\,\mathbf{G}^r_0(\epsilon)\,\mathbf{\Gamma}^Kf_K(\epsilon)\,
 \mathbf{G}^a_0(\epsilon)
\end{align}
where $\mathbf{G}^a_0(\epsilon)\equiv\left[\mathbf{G}^r_0(\epsilon)\right]^\dagger$.

Below we calculate the charge pumped through the junction by the optical pulse
\begin{equation}
\label{Qt}
Q(t) \equiv \int_0^t dt'\, \frac{I_L(t')-I_R(t')}{2} - I_0\, t
\end{equation}
where $I_{L,R}(t)$ are defined in Eq.(\ref{IK}), and $I_0$ is the steady-state current
\begin{equation}
 I_0\equiv \int\frac{d\epsilon}{2\pi}\,\mbox{Tr}
 \Big[\mathbf{\Gamma}^L\,\mathbf{G}^r_0(\epsilon)\,\mathbf{\Gamma}^R\,\mathbf{G}^a_0(\epsilon)\Big]
 \Big(f_L(\epsilon)-f_R(\epsilon)\Big)
\end{equation}

\section{\label{EM}Self-consistent electrodynamics}
The time evolution of electric, $\vec E$, and magnetic, $\vec H$, fields is considered according to the set of Maxwell's equations (written here in SI units)
\begin{subequations}
\label{Maxwell}
 \begin{eqnarray}
 \mu_0\frac{\partial\vec H(\vec r,t)}{\partial t}& = &-\vec\nabla\times\vec E(\vec r,t), \label{Faraday}
 \\
 \epsilon_0\frac{\partial \vec E(\vec r,t)}{\partial t}& = &\vec\nabla\times\vec H(\vec r,t)-\vec J(\vec r,t),  \label{Ampere}
 \end{eqnarray}
\end{subequations}
where $\mu_0$ and $\epsilon_0$ are the magnetic permiability and dielectric permittivity of 
the free space, respectively, and $\vec J(t)$ is the electric current density.
Note that magnetization is disregarded in Eqs.~(\ref{Faraday}) and (\ref{Ampere}),
since we assume both molecule and contacts to be non-magnetic.

A molecule located at site $m$ (${}\equiv\vec r_m$) and driven by local electric field $\vec E(\vec r_m,t)$, yields time-dependent response, which enters Ampere's law as a polarization current density
\begin{equation}
\label{JP}
 \vec J(\vec r_m,t) = \frac{\partial\vec P_m(t)}{\partial t}\delta(\vec r_m),
\end{equation}
where $\delta$ is the Dirac delta-function. The polarization depends on molecular characteristics through the molecular density matrix, which in turn is
affected by the local field. 
In our model two-level systems of the molecular bridge (\ref{HM}) are assumed to occupy 
sites of the FDTD grid.
Molecules contribute to the polarization at their site according to
\begin{equation}
\label{Pm}
 \vec P_m(t) = 2\,\mbox{Im}\left[\vec\mu_{mx,mg}\,G^{<}_{mg,mx}(t,t)\right]
\end{equation}

The resulting system of coupled differential equations, Eqs~(\ref{Maxwell})-(\ref{Ampere}), 
is solved simultaneously  with EOMs for the Green functions of the quantum system, 
Eqs.~(\ref{EOMGr})-(\ref{EOMGlt}).
The Maxwell's equations are discretized in time and space and propagated using the 
finite-difference time-domain approach (FDTD).\cite{Taflove:2005jj} 
We employ three-dimensional FDTD calculations utilizing home-build parallel FORTRAN-MPI 
codes on a local multi-processor cluster.\cite{plasmon-cluster} 
In spatial regions occupied by a plasmonic nanostructure (a bowtie antenna in our case)
we employ the auxiliary differential equation method to account for materials dispersion. 
The dielectric response of the metal is modeled using a standard Drude formulation with the set of parameters describing silver.\cite{FainbergPRB11,Sukharev2010}
The Green functions EOMs are propagated with the fourth order Runge-Kutta scheme.

Within described self-consistent model the local electric field $\vec E_m(t)\equiv\vec E(\vec r_m,t)$ in Eq.(\ref{HM}) driving a molecular junction is thus defined
by both SPP excitations in the contacts and the local molecular response. 
In the next section we show that the molecular contribution changes junction transport characteristics
drastically, and in general can not be ignored. 

\begin{figure}[t]
\centering\includegraphics[width=\linewidth]{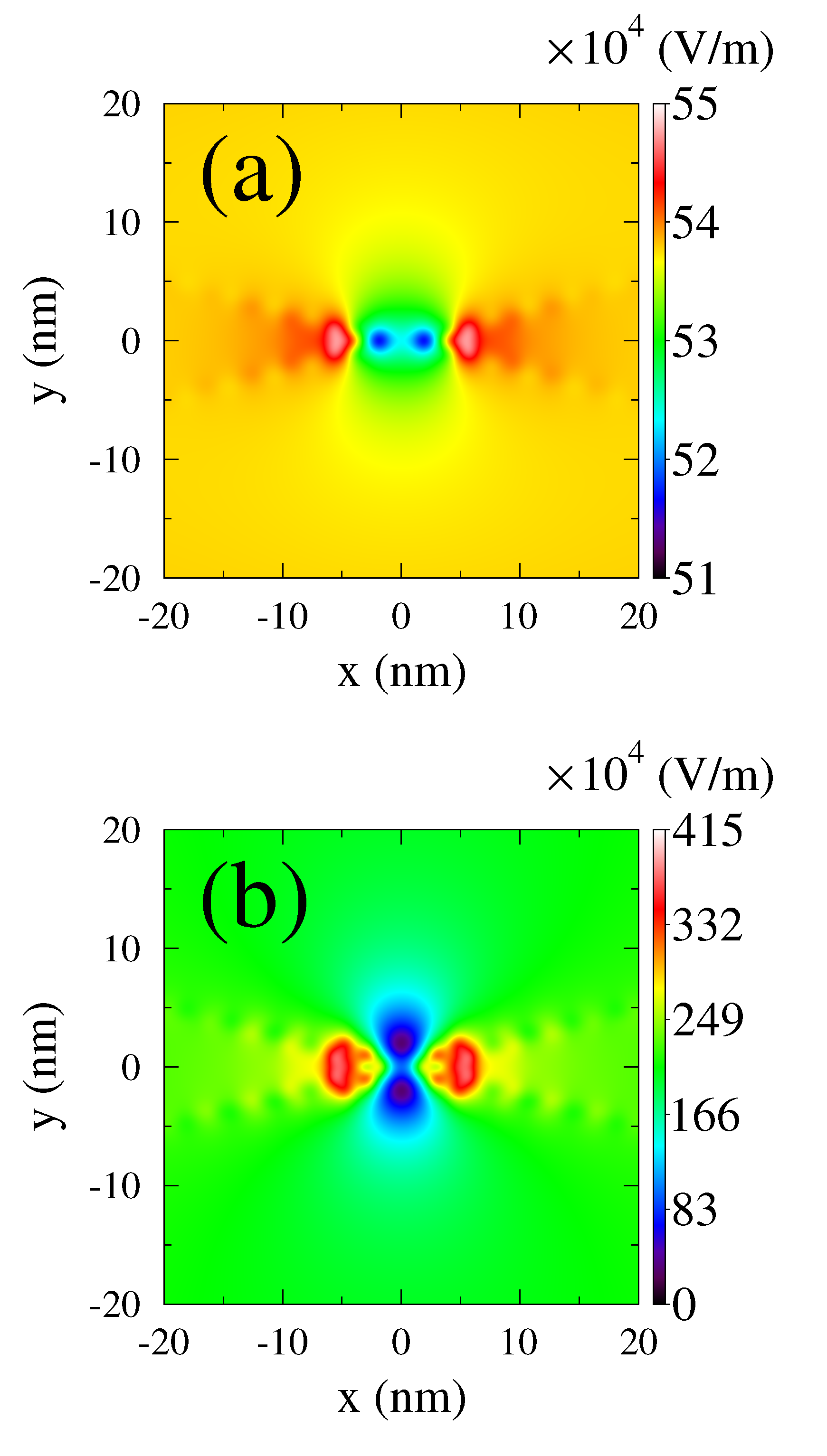}
\caption{\label{fig2}
(Color online) Map of the instantaneous electric field strength, 
$[E_x^{2} (\vec{r},t)+ E_y^{2} (\vec{r},t) + E_z^{2} (\vec{r},t)]^{1/2}$,  
at a distance of 10 nm from the molecule (the plane is parallel to $xy$) calculated
(a) without and (b) with the molecular response. 
The distribution is shown for $t=77.8$ fs and $81.7$ fs for (a) and (b) respectively.
See text for parameters.
}
\end{figure}
\section{\label{numres}Numerical results}
Here we present result of numerical simulations demonstrating the importance of a self-consistent
treatment of the local EM field dynamics. 
Previous studies considered the influence of an isolated molecule on plasmon 
transfer,\cite{LopataNeuhauserJCP09,NeuhauserJCP09,GaoNeuhauserJCP12} 
molecular features in 
absorption\cite{NordlanderNatMater10,NordlanderNL11,SukharevNitzanPRA11,WhiteFainbergGalperin12} 
and Raman\cite{JensenCR11,SchatzJPCA12,SchatzJPCC12} spectra of molecules
attached to nanoparticles. Below we discuss how molecular junctions and electron transport are influenced by a local EM field and vice versa in a self-consistent manner.

Unless otherwise specified parameters of the calculations are 
$T=300$~K, $\varepsilon_x=-\varepsilon_g=1$~eV, $t_x=t_g=0.05$~eV, 
$\mu_{gx}=32$~D, $\Gamma^L_{1g,1g}=\Gamma^R_{Dx,Dx}=0.1$~eV and
$\Gamma^L_{1x,1x}=\Gamma^R_{Dg,Dg}=0.01$~eV (other elements of the dissipation
matrix are zero). These parameters are chosen to represent a molecular junction with
a strong charge-transfer transition\cite{FainbergNitzanPRB07}, and are similar to our previous 
considerations.\cite{FainbergPRB11,Sukharev2010}
The Fermi energy is taken at the origin, $E_F=0$, and the bias is applied symmetrically,
$\mu_L=-\mu_R=V_{sd}/2$. 

Following Ref.~\onlinecite{FainbergPRB11}, the incoming incident
field is taken in the form of a chirped pulse
\begin{equation}
 E_{\text{inc}}(t) = \text{Re}\left[\mathcal{E}_0\exp\left(-\frac{(\delta^2-i\bar\mu^2)t^2}{2}-i\omega_0 t\right)\right]
\end{equation}
where $\mathcal{E}_0$ is the incident peak amplitude, $\omega_0$ is the incident frequency, and
$\delta^2\equiv2\tau_0^2/(\tau_0^4+4{\Phi''}^2(\omega_0))$ and 
$\bar\mu\equiv-4\Phi''(\omega_0)/(\tau_0^4+4{\Phi''}^2(\omega_0))$
are parameters describing the incident chirped pulse ($\tau_0$ is the characteristic time related
to the pulse duration). In the calculations below we use 
$\mathcal{E}_0=10^7$~V/m, $\omega_0=2$~eV, $\tau_0=11$~fs, and $\Phi''(\omega_0)=3000$~fs${}^2$.

Figure~\ref{fig2} shows instantaneous electric field strength distributions in a plane shifted by $z=10$~nm
parallel to $xy$ plane. The distribution is calculated for a junction
formed by bowtie antennas with single molecule ($D=1$) placed in the center of the gap.
Here $\varepsilon_x-\varepsilon_g=1.75$~eV, $\Gamma^L_{1g,1g}=\Gamma^R_{Dx,Dx}=0.01$~eV,
$\Gamma^L_{1x,1x}=\Gamma^R_{Dg,Dg}=0.001$~eV, and $V_{sd}=0$.
Fig.~\ref{fig2}a presents simulations without molecular response. Fig.~\ref{fig2}b shows results of a calculation where both SPP excitations in the contacts
 and molecular response are taken into account.  One can clearly see that even a single molecule drastically changes local electric field distribution.

\begin{figure}[t]
\centering\includegraphics[width=\linewidth]{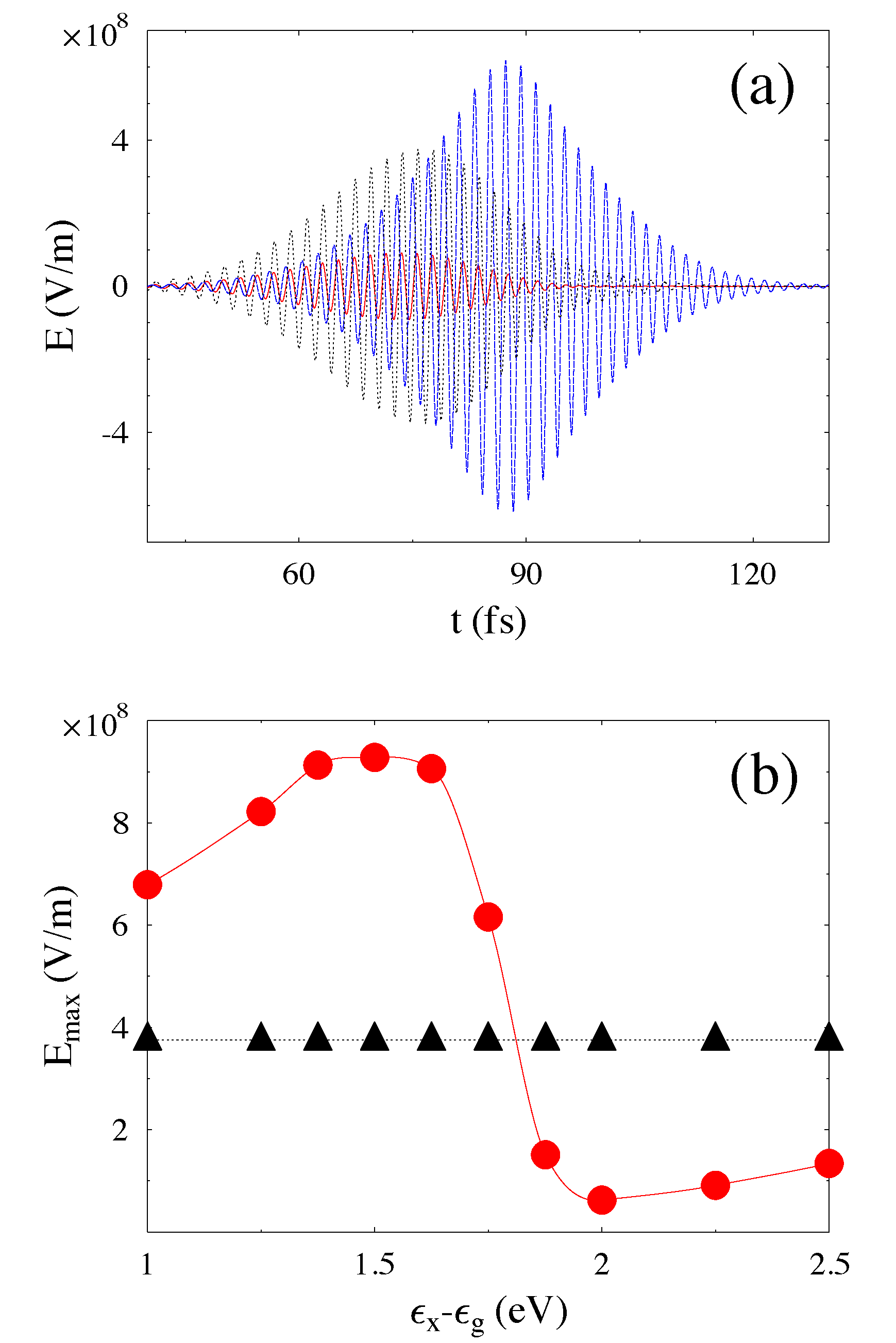}
\caption{\label{fig3}
(Color online) Local EM field at the molecular position.
(a) Pulse calculated without (dotted line, black) and with 
($\varepsilon_x-\varepsilon_g>\omega_0$ - solid line, red; 
$\varepsilon_x-\varepsilon_g<\omega_0$ - dashed line, blue) molecular response.
(b) Maximum local field during the pulse vs. molecular excitation energy calculated 
without (triangles, black) and with (circles, red) molecular response.
See text for parameters.
}
\end{figure}
\begin{figure}[t]
\centering\includegraphics[width=\linewidth]{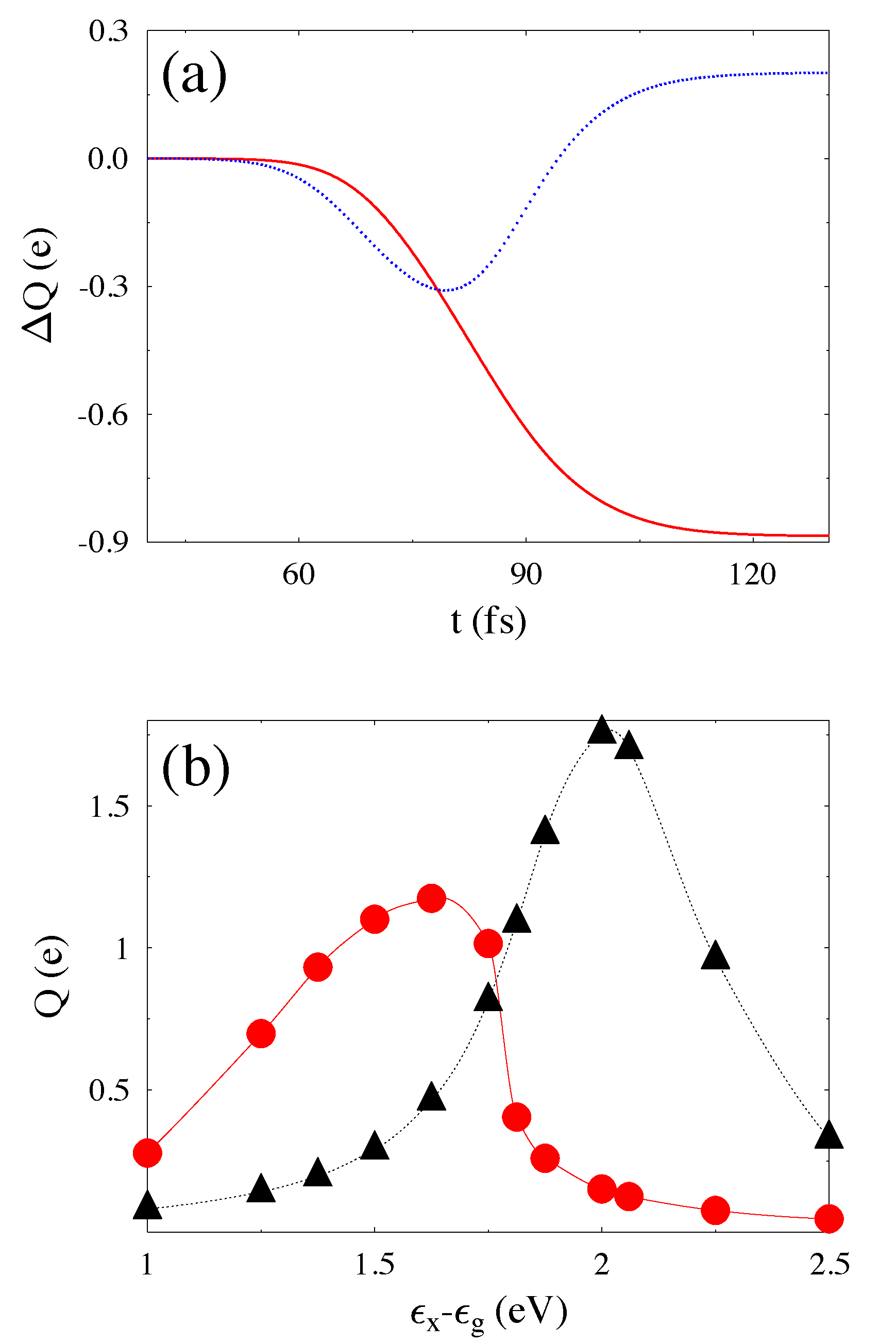}
\caption{\label{fig4}
(Color online) Charge pumped through the junction.
(a) Difference, $\Delta Q\equiv Q^{(sc)}-Q^{(nosc)}$, between results calculated with, $Q^{(sc)}$, 
and without, $Q^{(nosc)}$, molecular response vs. time for
$\varepsilon_x-\varepsilon_g>\omega_0$ (solid line, red) and 
$\varepsilon_x-\varepsilon_g<\omega_0$ (dotted line, blue).
(b) Total charge pumped during the pulse vs. molecular excitation energy calculated
without (triangles, black) and with (circles, red) molecular response.
See text for parameters.
}
\end{figure}

Sensitivity of the pulse temporal behavior to the molecular response is presented in Figure~\ref{fig3}a.
Here a local field affected by only SPP modes (dotted line) is compared to pulses calculated when molecular
response is taken into account. The latter may result in both enhancement (dashed line) or
quenching (solid line) of the local field depending on the ratio of the pulse frequency, $\omega_0$,
to the molecular excitation energy, $\varepsilon_x-\varepsilon_g$.
In particular, quenching is observed for the laser frequency being below the threshold
($\omega_0<\varepsilon_x-\varepsilon_g=2.25$~eV), while frequency above the threshold
($\omega_0>\varepsilon_x-\varepsilon_g=1.75$~eV) leads to enhancement of the field.
To understand this behavior we perform a simple analysis treating coupling to the
driving field as a perturbation, and neglecting the chirped character of the pulse. 
This leads to (see Appendix~\ref{appA})
\begin{align}
 \label{Papprox}
 P_1(t)\approx& -\mathcal{E}_0\cos(\omega_0 t)\,\lvert\mu_{gx}\rvert^2\int\frac{d\epsilon}{2\pi}\,
 \\
 &\left(\;\mbox{Im}\left[ G_{1g,1g}^{<}(\epsilon)\right]
 \frac{\epsilon-(\varepsilon_x-\omega_0)}
     {[\epsilon-(\varepsilon_x-\omega_0)]^2+[\Gamma_{1x,1x}/2]^2}
 \right. \nonumber \\
 &+\left.\mbox{Im}\left[ G_{1x,1x}^{<}(\epsilon)\right]
 \frac{\epsilon-(\varepsilon_g+\omega_0)}
     {[\epsilon-(\varepsilon_g+\omega_0)]^2+[\Gamma_{1g,1g}/2]^2}
 \right)
\nonumber
\end{align}
where $\mathbf{G}^{<}$ is the lesser projection of the Green function (\ref{G}). Taking into account that in the absence of the chirp 
$E_{\text{inc}}(t)=\mathcal{E}_0\cos(\omega_0 t)$ the first term in the right side of 
Eq.(\ref{Papprox}) suggests that for populated ground state, $G^{<}_{1g,1g}(\epsilon)\approx 1$,
the molecular polarization oscillates in phase with the field for 
$\omega_0<\varepsilon_x-\varepsilon_g$, and in anti-phase for 
$\omega_0>\varepsilon_x-\varepsilon_g$. Thus according to Eqs.~(\ref{Ampere}) and (\ref{Pm})
the molecular response quenches the field in the former case, and enhances it in the latter. 
Fig.~\ref{fig3}b illustrates this finding within the exact calculation showing the maximum of the total field
for different molecular excitation energies (circles) compared to the maximum of the
EM field obtained without molecular response (triangles). 
Note that contribution of the second term in the right side of Eq.(\ref{Papprox}) 
is exactly the opposite that of the first term, however since the calculations presented in Fig.~\ref{fig3} are performed at zero bias, the molecular excited state is initially empty, $G^{<}_{1x,1x}(t=0)\approx 0$.

\begin{figure}[t]
\centering\includegraphics[width=\linewidth]{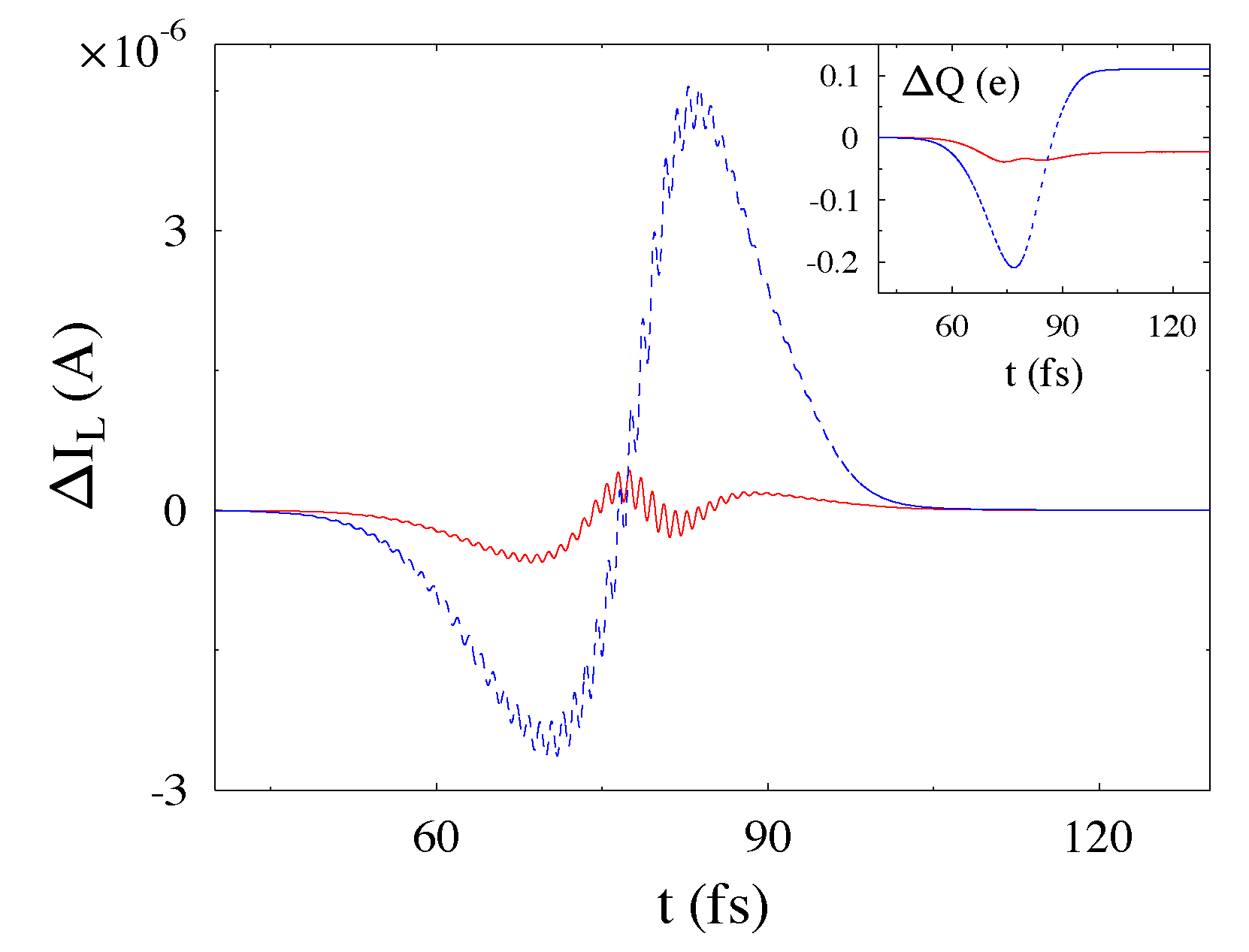}
\caption{\label{fig5}
(Color online) Current at the left interface as a function of time.
Shown are differences, $\Delta I_L\equiv I_{L}^{(sc)}-I_{L}^{(nosc)}$, between results
calculated with $I_{L}^{(sc)}$, and without, $I_{L}^{(nosc)}$, molecular response.
The calculations are performed for $V_{sd}=1.5$~V (dashed line, blue) and $2$~V (solid line, red).
Inset shows corresponding difference in charge pumped through the junction.
See text for parameters.
}
\end{figure}

While the local EM field cannot be measured directly, it is related to junction
characteristics (in particular, its transport properties) detectable in experiments.
Fig.~\ref{fig4}a demonstrates the difference in the temporal buildup of the charge pumped through
the junction, when the molecule is considered to be driven by the field obtained within the self-consistent model vs. model with only SPP excitations taken into account.
Initial dip in the charge buildup (see dotted line) is related to a time delay of the molecule induced pulse
for $\varepsilon_x-\varepsilon_g<\omega_0$
(compare solid and dashed lines to the dotted line in Fig.~\ref{fig3}a). 
The delay is caused by the chirped nature of the incoming pulse, with initial pulse frequency 
being lower than the molecular excitation energy, which results in 
suppression of the local field at the start of the pulse.
Eventually however the incoming frequency becomes higher than the molecular transition energy.
Corresponding enhancement of the local field leads to increase in the charge pumped through the junction.
Note that for $\varepsilon_x-\varepsilon_g>\omega_0$ no delay is observed, and the local field is 
quenched throughout the pulse. Correspondingly effectiveness of the charge pump is lower in this case
(see solid line in Fig.~\ref{fig4}a).

Figure~\ref{fig4}b shows 
the total charge pumped through the junction during the pulse at different molecular
excitation energies. Clearly, the most effective EM field obtained without the molecular response taken into account corresponds to 
the resonance situation, $\omega_0=\varepsilon_x-\varepsilon_g=2$~eV. When molecular
response is included in the model the situation is less straightforward. Since local field enhancement
is expected for low molecular excitation energies, $\omega_0>\varepsilon_x-\varepsilon_g$ 
(see Fig.~\ref{fig3}b), the peak in the pumped charge distribution is shifted to the left.
Note that the lower height of the shifted peak is related to the fact that, for a lower molecular gap,
part of optical scattering channels is blocked due to partial population of the broadened excited
and ground states of the molecule (see Ref.~\onlinecite{Sukharev2010} for detailed discussion). 

\begin{figure}[t]
\centering\includegraphics[width=\linewidth]{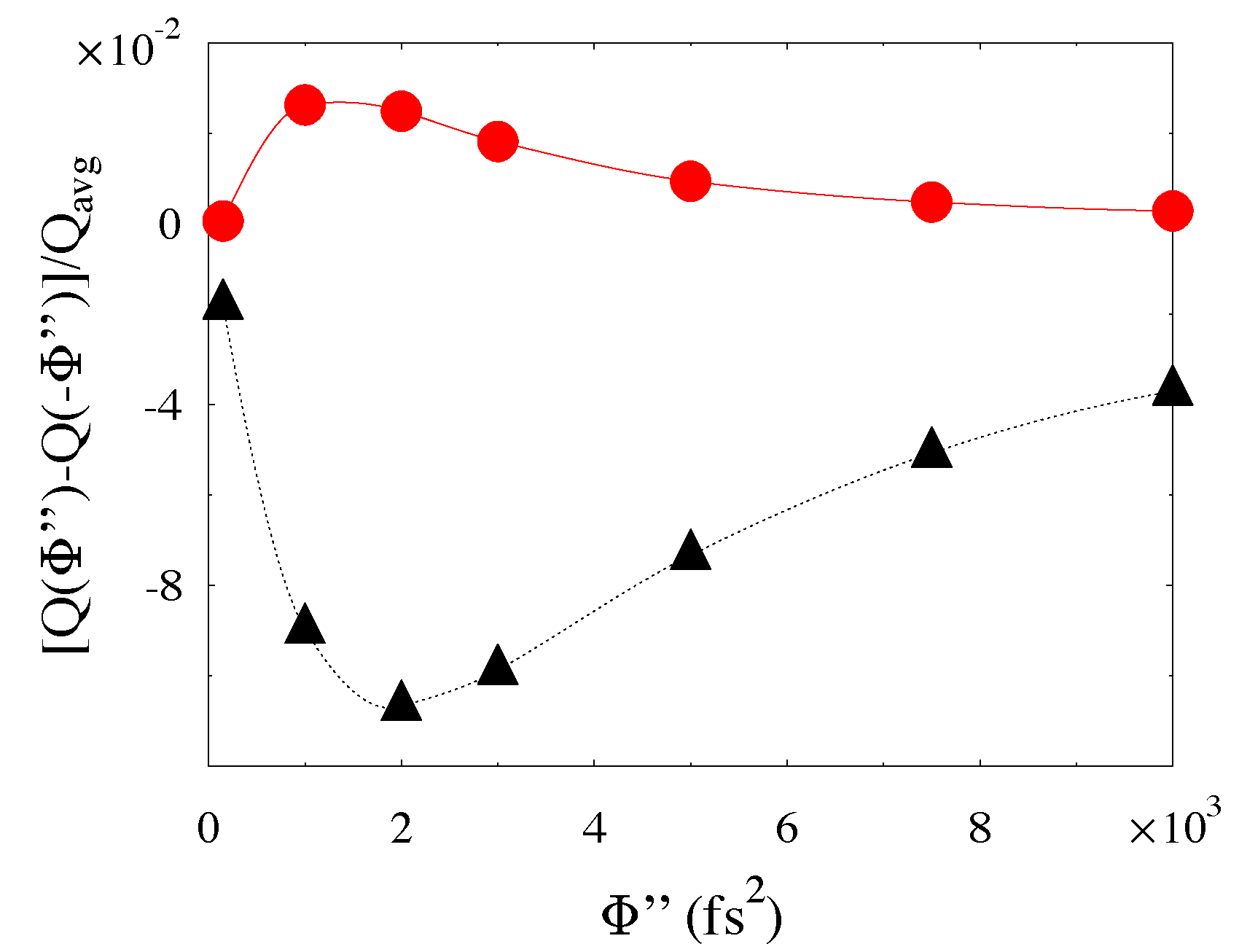}
\caption{\label{fig6}
(Color online) Asymmetry in the charge transfer between positively and negatively chirped 
incoming laser pulses, $Q(\Phi'')-Q(-\Phi'')$, normalized by their average, 
$Q_{avg}\equiv(Q(\Phi'')+Q(-\Phi''))/2$. 
Shown are results calculated without (triangles, black) and with (circles, red) 
the molecular response. See text for parameters.
}
\end{figure}

Note that the importance of molecular response depends also on bias across the junction. 
Indeed, since high bias, $V_{sd}>\varepsilon_x-\varepsilon_g$, may inject holes into 
the molecular ground state and electrons into the excited state, and since populating these
states has opposite consequences for the local field enhancement (see Eq.~(\ref{Papprox})
and the discussion following it), it is natural to expect that the molecular response is more 
important at low biases,
$V_{sd}<\varepsilon_x-\varepsilon_g$. 
Figure~\ref{fig5} illustrates this conclusion with results of our calculations within the self-consistent model. 
Here $\Gamma^L_{1x,1x}=\Gamma^R_{1g,1g}=0.05$~eV.
We observe that both difference in optically induced current 
and charge pumped through the junction (see inset) is almost negligible at high biases.
Similar reasoning indicates that the molecular response at strong incoming fields
will be less important also due to population of the excited molecular state induced 
by external pulse. 

\begin{figure}[t]
\centering\includegraphics[width=\linewidth]{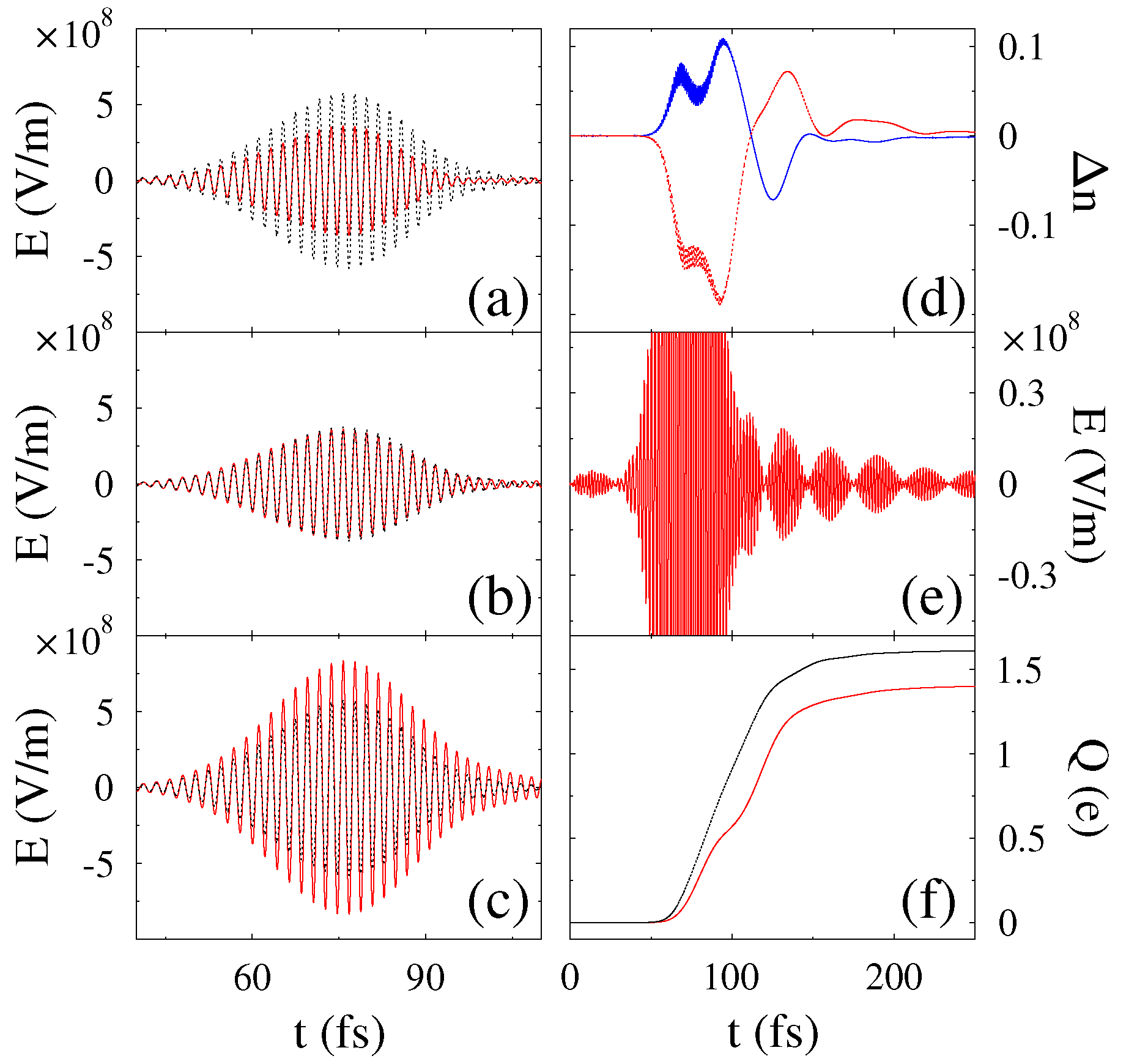}
\caption{\label{fig7}
(Color online) Effect of the self-consistent treatment on local field and level population 
in a $3$-sites molecular bridge ($D=3$) as functions of time. 
Shown are (a)-(c) local field calculated without (dotted line, black) and
with ($\varepsilon_x-\varepsilon_g>\omega_0$ - solid line, red) molecular response 
for the three molecular sites. Panel (d) shows the difference in population of the ground, 
$\Delta n_{1g}\equiv n_{1g}^{(sc)}-n_{1g}^{(nosc)}$ (solid line, blue)
and excited, $\Delta n_{1x}\equiv n_{1x}^{(sc)}-n_{1x}^{(nosc)}$ (dotted line, red) states for 
the first molecular site ($m=1$).
Panel (e) shows the scaled plot of the field on the central site ($m=2$) for a longer period of time. 
The  charge pumped through the $3$-sites molecular bridge vs. time is shown in panel (f).
See text for parameters.
}
\end{figure}

Asymmetry in the charge pumping relative to the sign of the chirp rate was discussed
in our recent publication (see Fig.~4 in Ref.~\onlinecite{FainbergPRB11}).
One of the reasons for the asymmetry is related to the time spent by the local pulse 
in the region of frequencies at and just below the resonance. This region provides
the main contribution to charge transfer (see discussion of Fig.~3 in ref.~\onlinecite{FainbergPRB11}). Since time spent in this region by the positively chirped pulse
is smaller than that by the pulse with equal negative chirp rate 
(the positively chirped local pulse is shorter), one expects to observe 
an asymmetry as represented by the result of calculations using local EM field influenced only by SPP modes
driving the junction (see curve with triangles in Fig.~\ref{fig6}). 
However as discussed above, it is this pre-resonance region where molecular response quenches
local field, thus diminishing (or even overturning) the asymmetry relative to the chirp rate
sign (see curve with circles in Fig.~\ref{fig6}). 

Finally, we consider a $3$-sites molecular bridge ($D=3$) to model the spacial nonlocality 
of molecular polarization. Calculations are done for 
$\varepsilon_{mx}-\varepsilon_{mg}=2.25$~eV, $\omega_0=2$~eV, and $V_{sd}=0$.
Panels (a)-(c) of Fig.~\ref{fig7} compare the pure plasmonic local field to the field calculated when the
molecular response is taken into account for the three sites of the bridge. 
Molecular polarization decreases the local field amplitude on the first site, (a), and
enhances it on the rightmost site, (c). The field at the middle site, (b), does not change.
The effect can be understood following the discussion similar to that of Fig.~\ref{fig3}. 
We find that for $\varepsilon_x-\varepsilon_g>\omega_0$
increase in population in the ground (decrease in the excited) levels of the molecular sites
quenches the local field. Change in the populations of the leftmost site, panel (a), resulting from self-consistent
treatment is shown in Fig.~\ref{fig7}d. We see that these changes are in 
agreement with the corresponding change in the local field. Similar considerations are also hold
for panels (b) and (c) (corresponding level population are not shown).

Self-consistently calculated electric field on a site in the bridge shows a visible beat at large timescale(see Fig.~\ref{fig7}e). This behavior is related to the Rabi frequency due to the intersite coupling, $t_s$.

Finally, Fig~\ref{fig7}f shows charge transferred through the $3$-site junction as function of time.
Decrease in the effectiveness of the pump is related to quenching of the local field on
the first site of the bridge, where strong coupling to the left contact yields quick resupply of
the ground level population. Decreased efficiency in pumping the charge between
ground and excited levels at this site is the reason for the overall change in the effectiveness
of the pump. 

\section{\label{conclude}Conclusion}
We consider a simple model of a molecular junction driven by external chirped laser pulses. 
The molecule is represented by a bridge of $D$ two-level systems. 
The contacts geometry are taken in the form of a bowtie antenna.
The FDTD technique is used to calculate the local field in the junction
resulting from SPP excitations in the contacts. Simultaneously we solve 
time-dependent nonequilibrum Green functions equations of motion to take into account
the molecular contribution to the local field formation. 

Note that many works on driven transport assume pure incident field to be a driving force
acting on the molecule. In our recent publications\cite{Sukharev2010,FainbergPRB11} 
we considered effects of local field formation due to SPP excitations in the contacts 
on junction characteristics under external optical pumping. Here we make one more step
by taking into account also the molecular response in the driving local field dynamics. 
Within a reasonable range of parameters
we demonstrate that the latter is crucial for proper description of the junction transport.
We compare our results with previously published predictions, and show that 
the molecular contribution may lead to measurable differences (both quantitative and qualitative)
in characteristics of junctions. This contribution is especially important at low biases 
and relatively weak external fields in the presence of a strong molecular transition dipole. 
In particular, we show that for laser frequencies
shorter (higher) than the molecular excitation energy the local SPP field is usually quenched 
(enhanced) by molecular response. 

Extension of the approach to realistic {\em ab initio}
calculations, taking into account time-dependent bias, and formulating a methodology for calculations in the language of molecular states are the goals for future research. 


\begin{acknowledgments}
M.G. gratefully acknowledges support
by the NSF (Grant No. CHE-1057930) and the BSF (Grant No. 2008282).
\end{acknowledgments}


\appendix
\section{\label{appA}Derivation of Eq.(\ref{Papprox})}
To understand trends observed in the exact calculations based on Eqs.~(\ref{EOMGr})-(\ref{EOMGlt}) and (\ref{Faraday})-(\ref{Ampere}), 
here we employ a simple consideration and derive an approximate expression for the molecular
polarization, Eq.(\ref{Pm}), given in Eq.(\ref{Papprox}). 
For simplicity we assume that only one projection of the molecular dipole is non-zero,
and consider a single molecule bridge ($D=1$). Then the molecular polarization is
\begin{equation}
 \label{appP1}
 P_1(t) = -2\mbox{Im}\left[\mu_{xg}\,G_{1g,1x}^{<}(t,t)\right]
\end{equation}
Assuming the dissipation matrix, Eq.(\ref{GammaK}), is diagonal the lesser projection
of the Green function in Eq.(\ref{appP1}) is given by the Keldysh equation of the form
\begin{align}
\label{appGlt}
 G_{1g,1x}^{<}(t,t)=&\sum_{s=g,x}\int_{-\infty}^t dt_1\int_{-\infty}^t dt_2\,
 G_{1g,1s}^{r}(t,t_1)\,
 \\ &\times
 \Sigma^{<}_{1s,1s}(t_1-t_2)\,G_{1g,1s}^{r}(t_2,t)
 \nonumber
\end{align}
where 
\begin{equation}
\label{appSlt}
\Sigma^{<}_{1s,1s}(t_1-t_2)=i \sum_{K=L,R}\int \frac{d\epsilon}{2\pi}\, f_K(\epsilon)\Gamma^K_{1s,1s} 
e^{-i\epsilon(t_1-t_2)}
\end{equation}
is the lesser self-energy due to coupling to the contacts.

We start by neglecting a chirp of the incoming field
\begin{equation}
\label{appEinc}
 E_{inc}(t)=\mathcal{E}_0\cos(\omega_0 t)
\end{equation}
and treat interaction between molecule and incoming field 
\begin{equation}
\label{appV}
V_{ss'}(t)\equiv -\delta_{s',\bar s}\,\mu_{s\bar s}\, E_{inc}(t)
\end{equation}
within the first order of perturbation theory. Here $\bar s$ indicates state opposite to $s$,
i.e. for $s=g$ $\bar s=x$.

Within the approximations the retarded Green function in Eq.(\ref{appGlt}) 
can be expressed as (similar expression can be written for the advanced projection)
\begin{align}
\label{appGr}
&G^r_{1s,1s'}(t,t')\approx\delta_{s,s'}G^{(0)r}_{1s,1s}(t-t')
\\
&+\sum_{m,n=g,x}\int_{-\infty}^{+\infty}dt''\,
G^{(0)r}_{1s,1s}(t-t'')\,V_{ss'}(t'')\,G^{(0)r}_{1s',1s'}(t''-t')
\nonumber
\end{align}
Here $\mathbf{G}^{(0)r}$ is the retarded projection of the Green functions (\ref{G}) in the absence
of external field
\begin{equation}
\label{appG0r}
 G^{(0)r}_{1s,1s}(t-t')=-i\theta(t-t')e^{-i(\varepsilon_s-i\Gamma_{1s,1s}/2)(t-t')}
\end{equation}
and $\theta(\ldots)$ is the Heaviside step function.

\begin{widetext}
Utilizing (\ref{appEinc})-(\ref{appG0r}) in (\ref{appP1})-(\ref{appSlt}) leads to
\begin{align}
 \label{appP1approx}
 &P_1(t)\approx -\mathcal{E}_0\,\lvert\mu_{gx}\rvert^2\int\frac{d\epsilon}{2\pi}\,
 \left(\;\mbox{Im}\left[ G_{1g,1g}^{(0)<}(\epsilon)\right]
 \frac{[\epsilon-(\varepsilon_x-\omega_0)]\cos(\omega_0 t)-[\Gamma_{1x,1x}/2]\sin(\omega_0 t)}
     {[\epsilon-(\varepsilon_x-\omega_0)]^2+[\Gamma_{1x,1x}/2]^2}
 \right. \\ &\qquad\qquad\qquad\qquad\qquad\quad
 +\left.\mbox{Im}\left[ G_{1x,1x}^{(0)<}(\epsilon)\right]
 \frac{[\epsilon-(\varepsilon_g+\omega_0)]\cos(\omega_0 t)+[\Gamma_{1g,1g}/2]\sin(\omega_0 t)}
     {[\epsilon-(\varepsilon_g+\omega_0)]^2+[\Gamma_{1g,1g}/2]^2}
 \right)
\nonumber
\end{align}
\end{widetext}
where we have used the Keldysh equation for the steady state situation
\begin{equation}
G_{1s,1s}^{(0)<}(\epsilon) =\frac{\sum_{K=L,R}if_K(\epsilon)\Gamma^K_{1s,1s}}
 {[\epsilon-\varepsilon_s]^2+[\Gamma_{1s,1s}/2]^2}
\end{equation}
Assuming that detuning is much bigger than levels broadenings,
$\lvert\omega_0-(\varepsilon_x-\varepsilon_g)\rvert\gg\Gamma_{1s,1s}$ ($s=g,x$),
the term with $\sin(\omega_0 t)$ in (\ref{appP1approx}) can be ignored. 
Finally, dressing the Green functions in Eq.~(\ref{appP1approx}), i.e. taking into account diagrams 
related to population redistribution in the molecule due to presence of the driving field, 
leads to Eq.(\ref{Papprox}).


\end{document}